\documentclass{ifacconf}

\usepackage{empheq}
\usepackage{changes}
\usepackage{cite}
\usepackage{graphicx}
\graphicspath{{Figs/}{../pdf/}{../jpeg/}}
\DeclareGraphicsExtensions{.pdf,.jpeg,.png}
\usepackage{lipsum}
\usepackage{algorithm}
\usepackage{array}
\usepackage{fixltx2e}
\usepackage{stfloats}
\usepackage{url}
\usepackage{booktabs}
\usepackage{enumerate}
\usepackage{setspace}

\usepackage{bm}
\usepackage{amsmath}
\usepackage{amssymb}
\usepackage{mathrsfs}
\newtheorem{theorem}{Theorem}[section]
\newtheorem{lemma}[theorem]{Lemma}

\definecolor{CBLUE}{RGB}{0,114,189}
\definecolor{CRED}{RGB}{217,83,25}
\definecolor{CYELLOW}{RGB}{237,177,32}
\definecolor{CPURPLE}{RGB}{126,47,142}

\usepackage{natbib}
\begin{document}
\begin{frontmatter}

\title{Geometric Decentralized Stability Certificate for Power Systems Based on Projecting DW Shells} 
% Title, preferably not more than 10 words.

\thanks[footnoteinfo]{This work was supported by the National Natural Science Foundation of China (U24B6008 and U22B6008).}

\author[First]{Linbin Huang,} 
\author[First]{Liangxiao Luo,} 
\author[First]{Ruohan Leng,} 
\author[First]{Huanhai Xin,}
\author[Second]{Dan Wang,}
\author[Third]{and Florian D\"orfler}

\address[First]{College of Electrical Engineering, Zhejiang University, China.\\ (e-mails: \{hlinbin, luolx, lengruohan, xinhh\}@zju.edu.cn).} 
\address[Second]{School of Robotics and Automation, Nanjing University, China. (e-mail: danwang@nju.edu.cn).} 
\address[Third]{Department of Information Technology and Electrical Engineering, ETH Zürich, Switzerland. (e-mail: dorfler@ethz.ch).}

\begin{abstract}                
The development of decentralized stability conditions has gained considerable attention due to the need to analyze multi-agent network systems, such as heterogeneous multi-converter power systems. A recent advance is the application of the small-phase theorem, which extends the passivity theory. However, it requires the transfer function matrix to be sectorial, which may not hold in some frequency range and will result in conservativeness. To address this issue, this paper proposes a geometric decentralized stability condition based on Davis-Wielandt (DW) shell and its projections. Our approach provides a geometric interpretation of the small-gain and small-phase theorems and enables decentralized stability analysis of power systems. It serves as a visualization method to understand the closed-loop interactions and assess the stability of large-scale network systems in a scalable and modular manner.\vspace{-2mm}
\end{abstract}


\vspace{-2mm}

\end{frontmatter}

	% \begin{IEEEkeywords}
	% Decentralized stability conditions, power converters, power systems, DW shells,  small signal stability.
	% \end{IEEEkeywords}
	
	\section{Introduction}\label{sec:intro}

    \vspace{-1mm}
	
	Modern power systems feature the large-scale integration of heterogeneous power electronics converters~\citep{milano2018foundations,xin2022many}. Such heterogeneity comes from the various types of devices (wind, solar, energy storage systems, etc.), the various deployed control schemes such as grid-following control and grid-forming control, as well as various vendors. The stability analysis of heterogeneous multi-converter systems is an intricate task, as it involves high-dimensional dynamics and complex closed-loop interactions. Hence, modular stability analysis approaches based on decentralized conditions are sought.

    Recent works on this topic include the application of the small-phase theorem, combined with the small-gain theorem, to perform decentralized (small-signal) stability analysis of power electronics-dominated power systems~\citep{Huang2024}. The small-phase theorem builds upon the concept of matrix phases, which is defined using the numerical range~\citep{chen2024phase}. This approach has shown its capability in analyzing highly heterogeneous multi-converter systems in a decentralized manner, that is, if each converter satisfies certain conditions, then the interconnected multi-converter system is guaranteed to be stable. The decentralized stability analysis will be an important cornerstone for future power systems, as it helps reveal how heterogeneous converters interact with the power network and thereby affect the system stability.

    However, the calculation of matrix phases requires that the matrix is sectorial, which may be conservative under certain scenarios. To tackle this problem, this paper develops a geometric decentralized stability condition based on the concept of Davis-Wielandt (DW) shell. We introduce the conditions of DW shell separation, numerical range separation, \(x\)-\(z\) graph separation, small gain, and small phase, and further provide self-contained proofs to bridge these conditions. Building on this, we propose a geometric decentralized stability condition by combining the information of matrix gain, phase, and \(x\)-\(z\) graph (the projection of DW shell onto the \(x\)-\(z\) plane). Our method enables scalable and modular stability analysis of large-scale network systems and provides a geometric interpretation of the closed-loop stability. The effectiveness of the method is validated in power system applications.

    \vspace{-0.5mm}
    \section{Preliminaries}\label{sec:II}
    \vspace{-2mm}

    \begin{figure}[!t]
	\centering
	\includegraphics[width=2.8in]{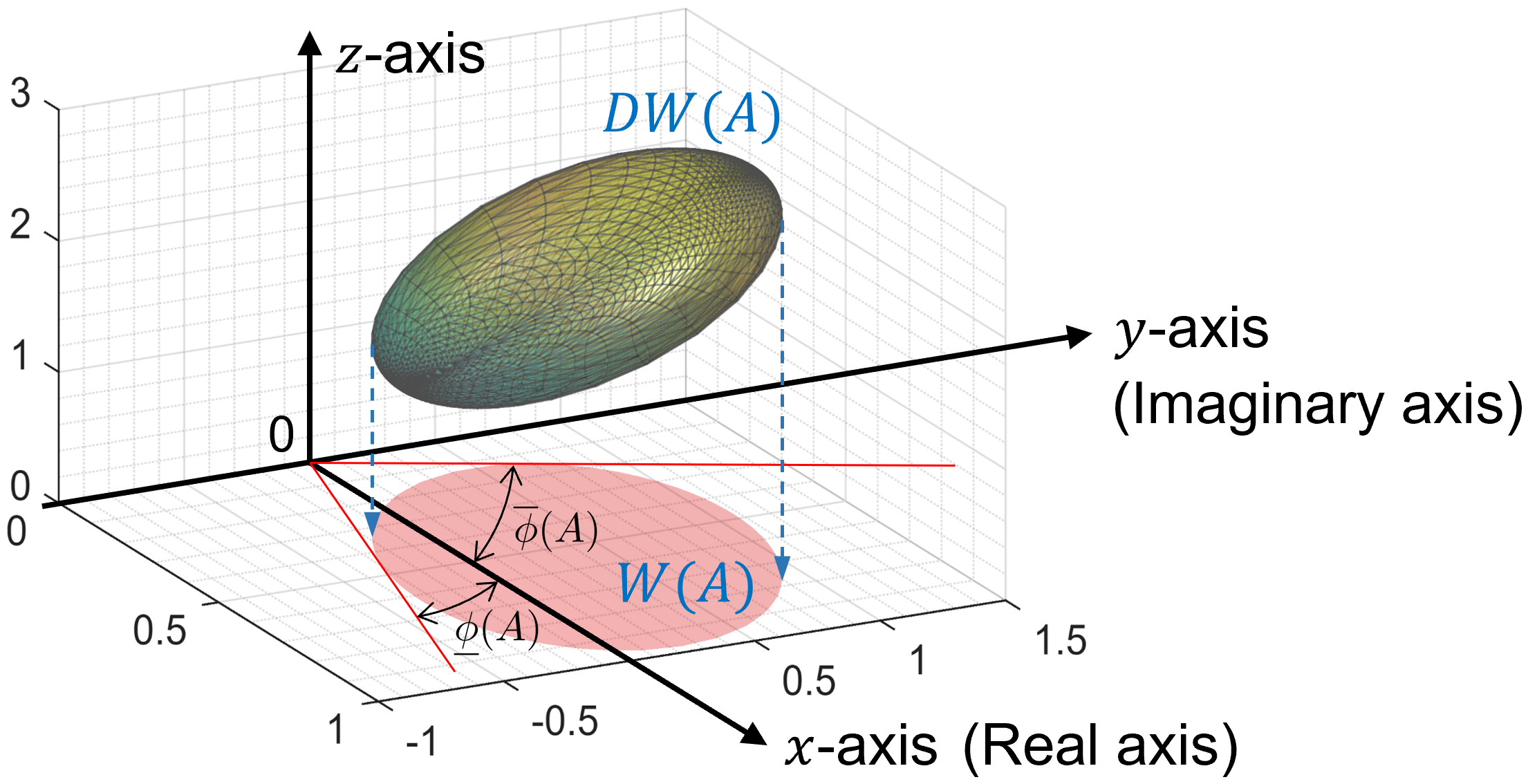}
	\vspace{-1mm}
	%\DeclareGraphicsExtensions.
	\caption{Illustration of the DW shell $DW(A)$, numerical range $W(A)$, and the phases $\underline\phi(A)$, $\overline{\phi}(A)$ of a sectorial matrix $A$. The (2D) numerical range is the projection of the (3D) DW shell onto the $x$-$y$ plane.}
	\vspace{-1mm}
	\label{Fig_DW_shell}
    \end{figure}

    The \textbf{numerical range} of a complex matrix $A \in \mathbb{C}^{n \times n}$ is 
\begin{equation}\label{eq:WA}
W(A)=\{x^*Ax: x\in \mathbb{C}^n, \|x\|=1\},
\end{equation}
    which is a convex subset of $\mathbb{C}$ and contains the eigenvalues of $A$. Here, $x^*$ denotes the conjugate transpose of $x$, and $\|x\|$ denotes the two-norm of $x$. If $0\notin W(A)$, then $A$ is said to be \textit{sectorial}. 
    For a sectorial $A$, there exists a nonsingular matrix $T$ and a diagonal unitary matrix $D$ such that 
$A=T^*DT$, referred to as the sectorial decomposition.
Here $D$ is unique up to permutation and has all diagonal entries lying on an arc of the unit circle with length smaller than $\pi$. 
Then, the \textit{phases} of $A \in\mathbb{C}^{n\times n}$, denoted by 
\begin{equation}
    \overline\phi(A)=\phi_1(A) \geq \dots \geq \phi_n(A)=\underline\phi(A), 
    \tag{\bf Phases}
\end{equation}
are defined as the phases of the eigenvalues of $D$ so that $\overline\phi(A)-\underline\phi(A)<\pi$~\citep{chen2024phase,wang2020phases}. 
    The smallest phase $\underline\phi(A)$ and the largest phase $\overline{\phi}(A)$ of a sectorial matrix $A$ are the angles formed by the positive real axis and the two supporting rays of $W(A)$, as shown in Fig.~\ref{Fig_DW_shell}, and the other phases of $A$ lie in between.

    As a counterpart, the gains (magnitudes) of a complex matrix $A\in\mathbb{C}^{n\times n}$ can be defined by its singular values
\begin{equation}
    \overline{\sigma}(A) =\sigma_1(A) \geq \dots \geq \sigma_n(A)=\underline{\sigma}(A).
    \tag{\bf Gains}
\end{equation}

The above concepts of ``phases'' and ``gains'' describe the properties of a complex matrix from different perspectives, and lead to different stability criteria of interconnected systems, namely, the small-gain theorem and the small-phase theorem~\citep{zames2003input, chen2024phase}. In recent years, the concept of DW shell, tracing back to the seminal works by Davis and Wielandt in~\citep{wielandt1955eigenvalues} and~\citep{davis1968shell}, has gained renewed interest thanks to its ability to simultaneously describe the gain and phase features of a complex matrix~\citep{zhang2025phantom,lestas_DW,Li_DW}.
To be specific, the \textbf{DW shell} of a complex matrix $A \in \mathbb{C}^{n \times n}$ is
\begin{equation}\label{eq:DWA}
DW(A)=\{(x^*Ax,\|Ax\|^2): x\in \mathbb{C}^n, \|x\|=1\},
\end{equation}
which extends the numerical range~\eqref{eq:WA} from the 2D complex plane to the 3D space, where the extra $z$-axis reflects the matrix gains. 
Fig.~\ref{Fig_DW_shell} shows the DW shell of an exemplary matrix $A$, where its projection onto the $x$-$y$ plane is the numerical range of $A$. Moreover, the top and bottom points of $DW(A)$ (regarding the $z$-axis) correspond to the square of the largest and smallest singular values of $A$, respectively. We introduce some important properties of the DW shell of $A \in \mathbb{C}^{n \times n}$:

%\vspace{1.5mm}
\noindent a) $DW(A) = DW(U^*AU)$ under any unitary matrix $U$;

%\vspace{1.5mm}
\noindent b) for a block-diagonal matrix $A={\rm diag}(A_1,A_2,...,A_n)$ where $A_i$ is the $i$-th block, we have 
\[
DW(A) = {\rm conv}\{DW(A_1),DW(A_2),...,DW(A_n)\}\,,
\]
where ${\rm conv}\{\cdot\}$ denotes the convex hull;

%\vspace{1.5mm}
\noindent c) $DW(A) \subset \{(c = x+jy \in \mathbb{C},z \in \mathbb{R}): z \ge x^2+y^2 \}$, where $x$ and $y$ are the real part and imaginary part of $c$;

%\vspace{1.5mm}
\noindent d) if $DW(A) \cap DW(-B) = \varnothing$, then ${\rm det}(A+B) \ne 0$.
%\vspace{1.5mm}

The above properties of DW shell will be used in the stability analysis of this paper, and we refer to~\citep{zhang2025phantom} and~\citep{Li_DW} for more detailed properties and analysis on DW shell.

    \section{Modeling of Multi-Converter Systems}\label{sec:III}

    \vspace{-1.5mm}

    \begin{figure}[!t]
	\centering
	\includegraphics[width=3.0in]{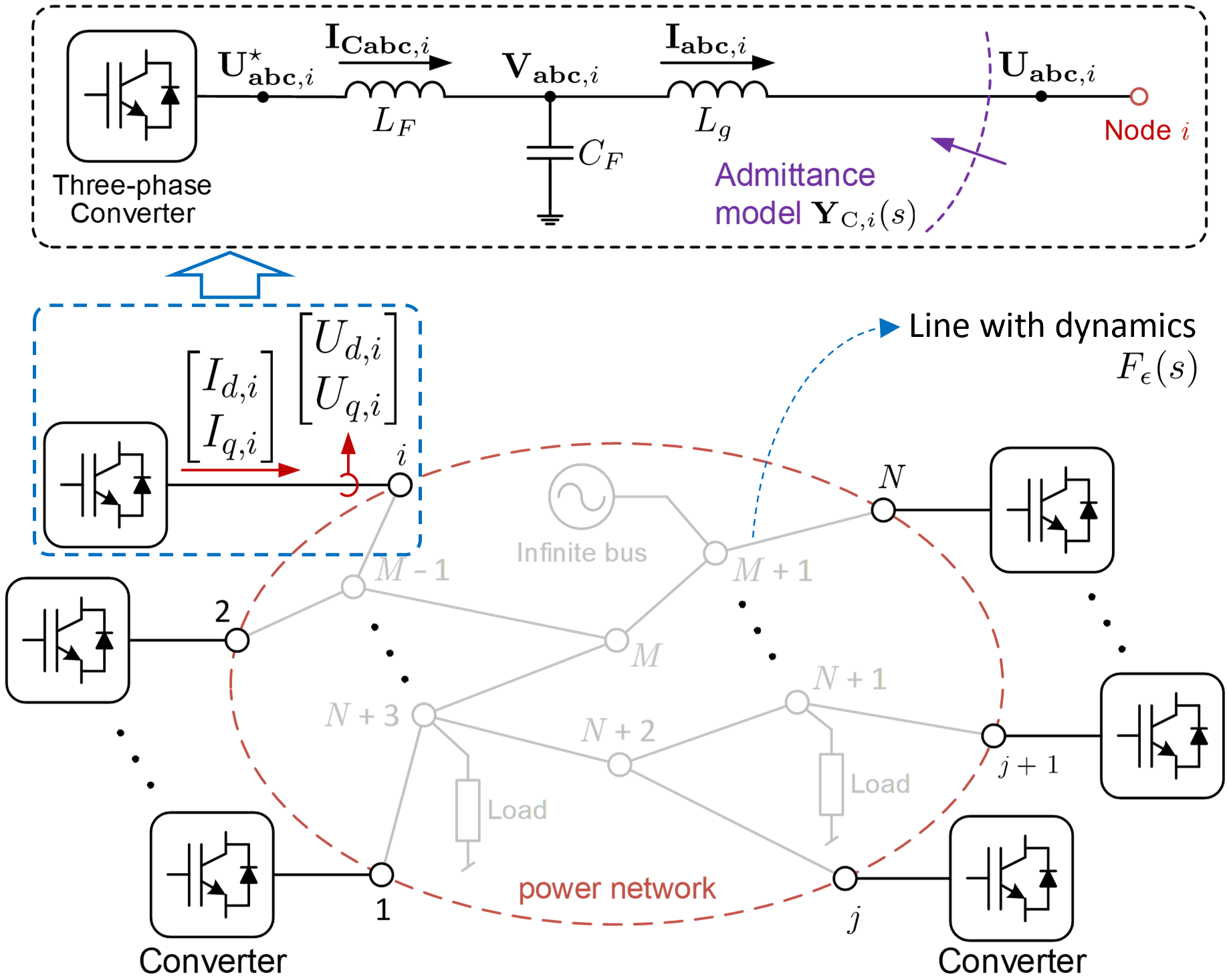}
	\vspace{-2mm}
	%\DeclareGraphicsExtensions.
	\caption{Illustration of a multi-converter system.}
	\vspace{1mm}
	\label{Fig_multiConv}
    \end{figure}

\begin{figure}[!t]
    \vspace{-2mm}
	\centering
	\includegraphics[width=3.1in]{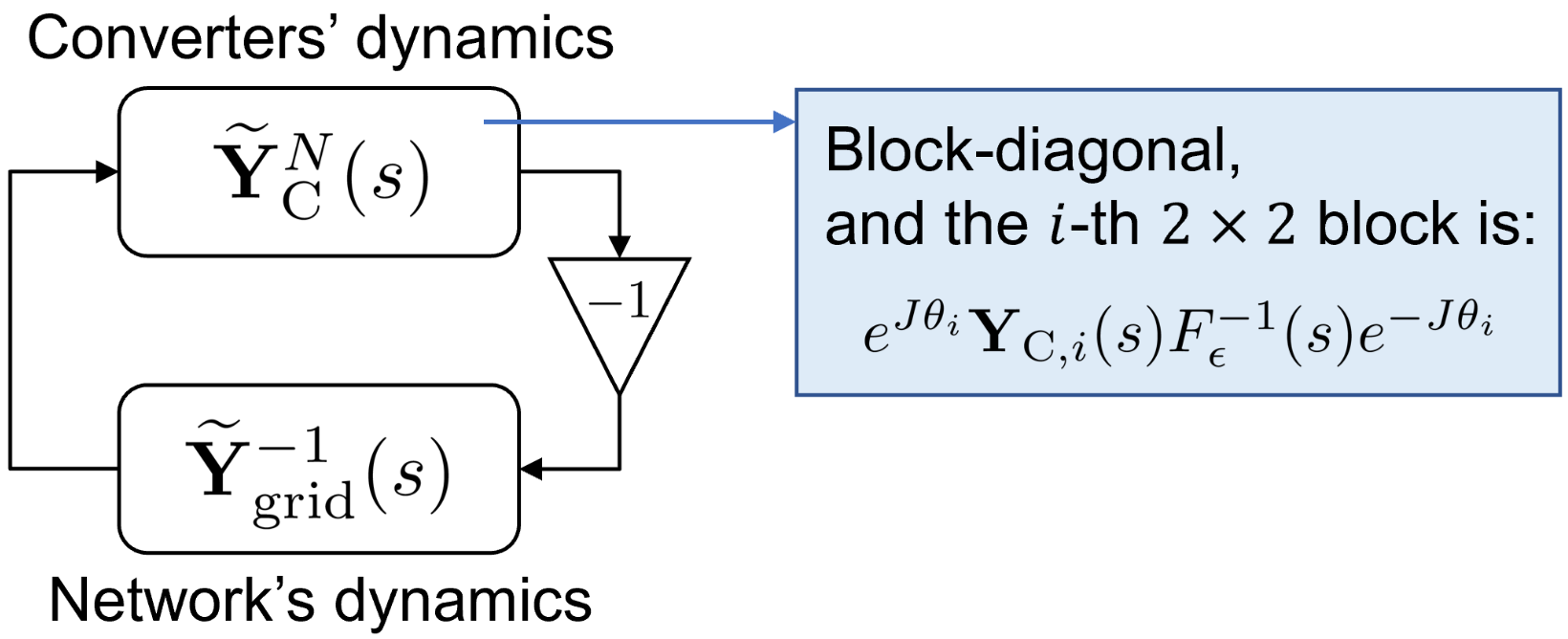}
	\vspace{-2mm}
	%\DeclareGraphicsExtensions.
	\caption{Closed-loop interaction of network and converters.}
	\vspace{1mm}
	\label{Fig_closed_loop}
    \end{figure}

    Consider a multi-converter power system as illustrated in Fig.~\ref{Fig_multiConv}, where $N$ heterogeneous converters are interconnected via the power network. The converters can be used to represent renewable energy generators, high-voltage DC stations, and so on, which are important components of modern power systems. As detailed in~\citep{Huang2024} and ~\citep{huang2020h}, such a multi-converter power system can be modeled as the closed-loop feedback system in Fig.~\ref{Fig_closed_loop}, and its characteristic equation is 
    \begin{equation}\label{eq:characteristic}
        {\rm det}(I+\widetilde{\bf Y}_{\rm grid}^{-1}(s)\widetilde{\bf Y}_{{\rm C}}^N(s) ) = 0\,,
    \end{equation}
    where ${\rm det}(A)$ denotes the determinant of matrix $A$. In this paper, we use the same notations as in~\citep{Huang2024}, where $\widetilde{\bf Y}_{\rm grid}(s)$ is the $2N \times 2N$ grid admittance (a transfer function matrix) capturing the power network dynamics, and $\widetilde{\bf Y}_{{\rm C}}^N(s)$ is a $2N \times 2N$ block-diagonal transfer function matrix (with $N$ $2 \times 2$ blocks) capturing the dynamics of the $N$ converters, which are stable in open loop.

    We assume that all the transmission lines in the power network have the same R/X ratio. In this case, according to the modeling procedure in~\citep{Huang2024}, the power network characteristics can be expressed as
    \begin{equation}\label{eq:grid_dy}
        \widetilde{\bf Y}_{\rm grid}(j\omega) = {\bf S}^{-\frac{1}{2}} {\bf B}_{\rm r} {\bf S}^{-\frac{1}{2}} \otimes I_2,
    \end{equation}
    where ${\bf S} \in \mathbb{R}^{N \times N}$ is the diagonal capacity ratio matrix of the converters, $\otimes$ denotes the Kronecker product, $I_2$ denotes the $2 \times 2$ identity matrix, and ${\bf B}_{\rm r}\in \mathbb{R}^{N \times N}$ is the Kron-reduced grounded Laplacian matrix of the power network obtained by eliminating the interior non-converter nodes~\citep{dong2018small}. We notice that in this special case (homogeneous R/X ratio), $\widetilde{\bf Y}_{\rm grid}(j\omega)$ becomes a positive-definite constant matrix. 

    The $i$-th block of $\widetilde{\bf Y}_{{\rm C}}^N(s)$ is denoted by $\widetilde{\bf Y}_{{\rm C},i}(s)$, which captures the dynamics of the \text{$i$-th} converter as
    \begin{equation}\label{eq:conv_dy}
        \widetilde{\bf Y}_{{\rm C},i}(s) := e^{J\theta_i} {\bf Y}_{{\rm C},i}(s){F}_\epsilon^{-1}(s) e^{-J\theta_i},\; i \in \{1,..., N\},
    \end{equation}
    where ${\bf Y}_{{\rm C},i}(s)$ is the original admittance model of the converter, which reflects how the converter's current responds to the terminal voltage perturbations, as indicated in Fig.~\ref{Fig_multiConv}, $e^{J\theta_i} = \begin{bmatrix} \cos \theta_i & -\sin \theta_i \\ \sin \theta_i & \cos \theta_i \end{bmatrix} \vspace{1mm}$ is a unitary matrix to handle the angle difference among the converters, and 
    \begin{equation}\label{eq:L_dy}
    {F}_\epsilon(s) = \begin{bmatrix} s/\omega_0 + {\epsilon} & -1 \\ 1 & s/\omega_0 + {\epsilon}
    \end{bmatrix}^{-1}
    \end{equation}
    represents the line dynamics under the same R/X ratio ${\epsilon}$. Originally, such dynamics should be included in the power network side, while here, we include them in the converter side so as to simplify the power network dynamics.

    The converters' dynamics interact with the power network in a closed-loop manner, as illustrated in Fig.~\ref{Fig_closed_loop}, with the characteristic equation given in~\eqref{eq:characteristic}. 
    The generalized Nyquist criterion (GNC) can be applied to assess the closed-loop stability, namely, the system is stable if and only if the Nyquist plot of ${\rm det}(I+\widetilde{\bf Y}_{\rm grid}^{-1}(s)\widetilde{\bf Y}_{{\rm C}}^N(s))$ does not encircle or pass through the origin. However, for large-scale systems with multiple converters, the GNC offers limited insight into closed-loop interactions and may encounter scalability issues. As a hope to overcome these limitations, we note that the system exhibits some special properties, namely, $\widetilde{\bf Y}_{{\rm C}}^N(s)$ is block-diagonal and $\widetilde{\bf Y}_{\rm grid}(s)$ is positive definite. 
    Leveraging these properties, we develop a decentralized stability condition in the next section to facilitate scalable and modular analysis of the system.

    \section{Geometric Stability Conditions}\label{sec:IV}
    \vspace{-2mm}
    This section develops geometric stability conditions. The principle of DW shell separation is introduced, which serves as a sufficient condition of GNC. Then, the geometric characteristics of DW shells are investigated, which give rise to a decentralized stability condition.

    \vspace{-1mm}
    \subsection{DW Shell Separation}
    \vspace{-1mm}
    Let \( G, H \in \mathcal{RH}_{\infty}^{m\times m} \), i.e., be real, rational, proper, and stable \( m\times m \) transfer function matrices ($H$ is invertible). Consider their standard feedback interconnection shown in Fig.~\ref{fdbk}. The closed-loop system is said to be stable if the \textit{Gang of Four matrix} describing all input-output maps
    \begin{align*}
    G(s)\#H(s)=\begin{bmatrix}
      (I + HG)^{-1} & (I + HG)^{-1}H\\
      G(I + HG)^{-1} & G(I + HG)^{-1}H
    \end{bmatrix}
    \end{align*}
    is stable, i.e., $G\#H \in \mathcal{RH}^{2m \times 2m}_\infty$.

    To analyze the closed-loop stability, we recall a sufficient version of the GNC, that is, \( G(s)\#H(s) \) is stable if, for each \( \omega \in [0,\infty) \), the following condition holds:
    \begin{equation}\label{eq:suff_GNC}
    \det\left(I + \tau G(j\omega) H(j\omega)\right) \neq 0, \quad \forall \tau \in (0, 1].
    \end{equation}
    
    The above condition ensures that the Nyquist plot of \( \det(I + G(j\omega)H(j\omega)) \) does not intersect with the origin or the negative real axis, indicating that it will not encircle or cross the origin, thereby confirming closed-loop stability.  
    We next show that the condition in~\eqref{eq:suff_GNC} can be guaranteed geometrically through the separation of the DW shells of \( G(j\omega) \) and \( H^{-1}(j\omega) \)~\citep{zhang2025phantom,lestas_DW}.
    \begin{lemma}[\bf{DW Shell Separation}]
    \label{prop:GeometricDSC}
    \,\\
    Consider $G, H \in\mathcal{RH}_{\infty}^{m\times m}$.
    The system \( G(s)\#H(s) \) is stable if, for each \( \omega \in [0,\infty) \),
    \begin{equation}\label{eq:DW_sep}
    DW(G(j\omega)) \cap DW(-\tfrac{1}{\tau} H^{-1}(j\omega)) = \varnothing , \, \forall \tau \in (0,1],
    \end{equation}
    that is, the DW shell of \( G(j\omega) \) is separated from the DW shell of \( H^{-1}(j\omega) \) scaled by \( -\frac{1}{\tau} \).
    \end{lemma}
    \begin{pf}
    Section~2 shows that \( DW(A) \cap DW(-B) = \varnothing \) implies \( \det(A + B) \neq 0 \). For $B$ nonsingular, we have \( \det(I + B^{-1}A) \neq 0 \), since \( \det(A + B) = \det(B) \times \det(I + B^{-1}A) \). Let \( A = G(j\omega) \) and \( B = -\frac{1}{\tau} H^{-1}(j\omega) \), we can derive~\eqref{eq:suff_GNC} from~\eqref{eq:DW_sep}, concluding that the system is stable. \unskip\nobreak\hfill $\square$
    \end{pf}

    \subsection{Geometric Separation Based on Projecting DW Shells}
    While the DW shell separation condition in Lemma~\ref{prop:GeometricDSC} provides a geometric interpretation for the closed-loop stability of \( G(s)\#H(s) \), it requires a full 3D characterization of the matrices. For easier visualization and interpretability, we develop several lower-dimensional geometric conditions as tractable alternatives, which are also indispensable for developing the decentralized stability condition.

    \setlength{\unitlength}{1mm}
    \begin{figure}[t!]
    \begin{center}
    \begin{picture}(50,20)
    \thicklines 
    \put(0,17){\vector(1,0){8}} \put(10,17){\circle{4}}
    \put(12,17){\vector(1,0){8}} \put(20,13){\framebox(10,8){$G(s)$}}
    \put(30,17){\line(1,0){10}} \put(40,17){\vector(0,-1){10}}
    \put(38,5){\vector(-1,0){8}} \put(40,5){\circle{4}}
    \put(50,5){\vector(-1,0){8}} \put(20,1){\framebox(10,8){$H(s)$}}
    \put(20,5){\line(-1,0){10}} \put(10,5){\vector(0,1){10}}
    \put(5,10){\makebox(5,5){$y_1$}} \put(40,10){\makebox(5,5){$y_2$}}
    \put(0,17){\makebox(5,5){$w_1$}} \put(45,0){\makebox(5,5){$w_2$}}
    \put(13,17){\makebox(5,5){$u_1$}} \put(32,0){\makebox(5,5){$u_2$}}
    \put(10,10){\makebox(6,10){$-$}}
    \end{picture}
    \vspace{-2mm}
    \caption{A standard closed-loop system $G(s)\#H(s)$.}
    \vspace{1mm}
    \label{fdbk}
    \end{center}
    \end{figure}

    \begin{lemma}[\bf{Numerical Range Separation}]
    \label{lem:NR}
    \,\\
    Consider $G, H \in \mathcal{RH}_{\infty}^{m\times m}$.
    The system \( G(s)\#H(s) \) is stable if, for each \( \omega \in [0,\infty) \),
    \begin{equation}\label{eq:NR_sep}
    W(G(j\omega)) \cap W\!\left(-\tfrac{1}{\tau}H^{-1}(j\omega)\right) = \varnothing, \,\forall \tau \in (0,1],
    \end{equation}
    that is, the numerical ranges of \( G(j\omega) \) and \( H^{-1}(j\omega) \) scaled by \( -\tfrac{1}{\tau} \) are separated in the complex plane (i.e., $x$-$y$ plane). Moreover, at any \( \omega \in [0,\infty) \), Eq.~\eqref{eq:NR_sep} implies Eq.~\eqref{eq:DW_sep}. 
    \end{lemma}
    
    \begin{pf}
    Note that \( W(G(j\omega)) \) and \( W(-\tfrac{1}{\tau}H^{-1}(j\omega)) \) 
    are the 2D projections of their DW shells onto the \( x\)-\(y \) plane. 
    When these numerical ranges are separated, the corresponding DW shells in the 3D space are also separated, 
    which leads to~\eqref{eq:DW_sep} and thus guarantees stability. \unskip\nobreak\hfill $\square$
    \end{pf}
    
    \begin{lemma}[\bf{\(\bm x\)-\(\bm z\) Graph Separation}]
    \label{lem:XZ}
    \,\\
    Consider $G, H \in \mathcal{RH}_{\infty}^{m\times m}$.
    The system \( G(s)\#H(s) \) is stable if, for each \( \omega \in [0,\infty) \),
    \begin{equation}\label{eq:XZ_sep}
    P(G(j\omega)) \cap P\!\left(-\tfrac{1}{\tau}H^{-1}(j\omega)\right) = \varnothing, \,\forall \tau \in (0,1],
    \end{equation}
    where \( P(A)= \{(\Re(x^*Ax), x^*Ax) : x \in \mathbb{C}^n, \|x\|=1\} \) is the projection of \( DW(A) \) onto the \(x\)-\(z\) plane, referred to as the \(x\)-\(z\) graph in this paper. Here $\Re(\cdot)$ denotes the real part. Moreover, at any \( \omega \in [0,\infty) \), Eq.~\eqref{eq:XZ_sep} implies Eq.~\eqref{eq:DW_sep}.
    \end{lemma}
    
    \begin{pf}
    The \(x\)-\(z\) graph is obtained by projecting the DW shell onto the \(x\)-\(z\) plane. If the \(x\)-\(z\) graphs of \( G(j\omega) \) and \( -\tfrac{1}{\tau}H^{-1}(j\omega) \) are separated, the corresponding DW shells in the 3D space are also separated, indicating~\eqref{eq:DW_sep} and hence ensuring stability by Lemma~\ref{prop:GeometricDSC}. \unskip\nobreak\hfill $\square$
    \end{pf}
    
    \begin{lemma}[\bf{Small Gain}]
    \label{lem:SG}
    Consider $G, H \in \mathcal{RH}_{\infty}^{m\times m}$.
    The system \( G(s)\#H(s) \) is stable if, for each \( \omega \in [0,\infty) \),
    \begin{equation}\label{eq:SG}
    \bar{\sigma}(G(j\omega))\,\bar{\sigma}(H(j\omega)) < 1.
    \end{equation}
    Moreover, at any \( \omega \in [0,\infty) \), Eq.~\eqref{eq:SG} implies Eq.~\eqref{eq:DW_sep}.
    \end{lemma}

    \vspace{0mm}
    \begin{pf}
    Eq.~\eqref{eq:SG} implies 
    \( \bar{\sigma}(G(j\omega)) < [\bar{\sigma}(H(j\omega))]^{-1} = \underline{\sigma}(H^{-1}(j\omega)) \).
    Hence, for any \( \tau \in (0,1] \),
    \begin{equation}\label{eq:SG_relation}
    \bar{\sigma}(G(j\omega)) < 
    \underline{\sigma}(H^{-1}(j\omega)) 
    \le \underline{\sigma}\!\left(-\tfrac{1}{\tau}H^{-1}(j\omega)\right).
    \end{equation}
    Since the maximum and minimum singular values correspond to the upper and lower bounds of a matrix’s DW shell along the \(z\)-axis, 
    inequality~\eqref{eq:SG_relation} implies that the DW shell of \(G(j\omega)\) lies entirely below that of \(-\tfrac{1}{\tau}H^{-1}(j\omega)\), i.e., they are separated. 
    By Lemma~\ref{prop:GeometricDSC}, this separation ensures the closed-loop stability of \( G(s)\#H(s) \). \unskip\nobreak\hfill $\square$
    \end{pf}

    \begin{lemma}[{\bf Small Phase}~\citep{chen2024phase}]
    \label{lem:SP}
    \hspace{-1mm}Consider \\ $G, H \in \mathcal{RH}_{\infty}^{m\times m}$.
    The system \( G(s)\#H(s) \) is stable if \( G(j\omega) \) and \( H(j\omega) \) are sectorial, and for each \( \omega \in [0,\infty) \), 
    \begin{equation}\label{eq:SP}
    \hspace{-1mm}
    \overline{\phi}(G(j\omega)) 
    \hspace{-0.4mm} + \hspace{-0.4mm}
    \overline{\phi}(H(j\omega)) \!<\! \pi,
    \underline{\phi}(G(j\omega)) 
    \hspace{-0.4mm} + \hspace{-0.4mm}
    \underline{\phi}(H(j\omega)) \!>\! -\pi.
    \end{equation}
    Moreover, at any \( \omega \in [0,\infty) \), Eq.~\eqref{eq:SP} implies Eq.~\eqref{eq:DW_sep}.
    \end{lemma}

    \begin{pf}
    Let $\gamma(H(j\omega)):=\frac{\overline\phi(H(j\omega))+\underline{\phi}(H(j\omega))}{2}$. Notice that 
    \begin{align*}
        \phi_i(H(j\omega))\!=\!\begin{cases}
            -\phi_{m-i+1}(-H^{-1}(j\omega)) \!+\! \pi & \text{if } \gamma(H(j\omega))\!>\!0\\
            -\phi_{m-i+1}(-H^{-1}(j\omega)) \!-\! \pi & \text{if } \gamma(H(j\omega))\!\leq\! 0
        \end{cases}
    \end{align*}\citep{wang2020phases}. By combining it with~\eqref{eq:SP} we obtain {\bf either}
    % \begin{equation}\label{eq:SP_substitute}
    % \left\{
    % \begin{aligned}
    % \overline{\phi}(G(j\omega))
    %   +\big[\pm\pi-\underline{\phi}(-H^{-1}(j\omega))\big] &< \pi,\\[2pt]
    % \underline{\phi}(G(j\omega))
    %   +\big[\pm\pi-\overline{\phi}(-H^{-1}(j\omega))\big] &> -\pi,
    % \end{aligned}
    % \right.
    % \end{equation}
    \( \overline{\phi}(G(j\omega)) < \underline{\phi}(-H^{-1}(j\omega)) \) 
    and 
    \( \underline{\phi}(G(j\omega)) > \overline{\phi}(-H^{-1}(j\omega))-2\pi \), {\bf or} \( \overline{\phi}(G(j\omega)) < \underline{\phi}(-H^{-1}(j\omega)) +2\pi \) 
    and 
    \( \underline{\phi}(G(j\omega)) > \overline{\phi}(-H^{-1}(j\omega)) \).   
    Both situation imply that the entire phase interval of \( G(j\omega) \) is separated from that of \( -H^{-1}(j\omega) \). 
    Since scaling by a positive real factor \( 1/\tau \) does not change the phase, 
    the same separation holds for \( -\tfrac{1}{\tau}H^{-1}(j\omega) \). 
    Hence, the numerical ranges \( W(G(j\omega)) \) and \( W(-\tfrac{1}{\tau}H^{-1}(j\omega)) \) are separated, and the corresponding DW shells are also separated. 
    By Lemma~\ref{prop:GeometricDSC}, such DW shell separation guarantees the closed-loop stability. \unskip\nobreak\hfill $\square$
    \end{pf}

    \subsection{Mixed Geometric Stability Condition}
    \vspace{-2mm}
    
    We note that Lemma~\ref{prop:GeometricDSC} formulates the DW shell separation in 3D space, 
    while Lemmas~\ref{lem:NR} and~\ref{lem:XZ} focus on its 2D projections onto the \(x\)-\(y\) and \(x\)-\(z\) planes, respectively. 
    Lemmas~\ref{lem:SG} and~\ref{lem:SP} further relate this geometric separation to the small-gain and small-phase theorems. 
    Collectively, these results offer multiple geometric perspectives with different levels of conservatism and computational burden. 
    To combine their respective advantages, we now present a mixed geometric stability condition that enables the use of different conditions across different frequency ranges.

    \begin{theorem}[{\bf Geometric Stability Condition}]
    \label{prop:FW}
    \,\\
    The system \( G(s)\#H(s) \) in Fig.~\ref{fdbk} is stable if for each \( \omega \in [0,\infty) \), 
    \textbf{either}
     \begin{enumerate}[1)]
        \item \label{cond:sg} the \textbf{small-gain condition}~\eqref{eq:SG} \textit{holds, \textbf{or}}
        \vspace{1mm}
    
        \item \label{cond:sp} the \textbf{small-phase condition}~\eqref{eq:SP} \textit{holds, \textbf{or}}
        \vspace{1mm}
    
        \item \label{cond:nr} the \textbf{numerical range separation}~\eqref{eq:NR_sep} \textit{holds, \textbf{or}}
        \vspace{1mm}
    
        \item \label{cond:xz} the \textbf{\(\bm{x}\)-\(\bm{z}\) graph separation}~\eqref{eq:XZ_sep} \textit{holds, \textbf{or}}
        \vspace{1mm}
    
        \item \label{cond:dw} the \textbf{DW shell separation}~\eqref{eq:DW_sep} \textit{holds, \textbf{or}}
        \vspace{1mm}

        \item the sufficient version of GNC~\eqref{eq:suff_GNC}
        \textit{holds.}
    \end{enumerate}
    \end{theorem}

    \begin{pf}
%conditions~\ref{cond:sg}~$\sim$~\ref{cond:xz} all imply the DW shell separation in condition~\ref{cond:dw}.  
    According to Lemmas~\ref{lem:SG}, \ref{lem:SP}, \ref{lem:NR}, and \ref{lem:XZ},
    for certain \( \omega \in [0,\infty) \), if one of the conditions~\ref{cond:sg}~$\sim$~\ref{cond:dw} is satisfied, the DW shells are separated, i.e., 
    \( DW(G(j\omega)) \cap DW(-\tfrac{1}{\tau}H^{-1}(j\omega)) = \varnothing \), implying that condition~6 holds at this $\omega$. Hence, if for each \( \omega \in [0,\infty) \), one of the conditions 1~$\sim$~6 holds,
    the system is stable according to GNC.  \unskip\nobreak\hfill $\square$
    %By Lemma~\ref{prop:GeometricDSC}, this ensures the stability of \( G(s)\#H(s) \). \unskip\nobreak\hfill $\square$
    \end{pf}

    Theorem~\ref{prop:FW} shows that different geometric stability conditions can be selectively used at different frequencies.  
    This enables a hierarchical analysis procedure: low-dimensional conditions, such as the small-gain and small-phase conditions, can be checked at the first step;  
    if they are not satisfied in certain frequency ranges, 2D geometric separation conditions ($x$-$z$ graph separation and numerical range separation) can then be examined;  
    lastly, if none of the above conditions are satisfied in certain frequency range, 3D DW shell separation condition can be checked, which is less conservative but requires higher computational effort.
    This strategy reduces the computational effort of stability analysis and serves as the basis for developing the geometric decentralized stability condition.

    \subsection{Geometric Decentralized Stability Condition}
    \vspace{-2mm}

    Based on the above conditions, we further exploit how special properties of $G(s)$ and $H(s)$ enable decentralized stability analysis. As discussed in Section~3, the multi-converter power system, modeled by the closed-loop interaction in Fig.~\ref{Fig_closed_loop}, exhibits two structural properties, namely, $\widetilde{\bf Y}_{{\rm C}}^N(s)$ is block-diagonal and $\widetilde{\bf Y}_{\rm grid}(s)$ is positive definite. To demonstrate the generality of our result, we consider $G(s) = \widetilde{\bf Y}_{{\rm C}}^N(s)$ exhibiting a block-diagonal structure, i.e.,
    \begin{equation}\label{eq:diag}
    G(s) = {\rm diag}\{ G_1(s), G_2(s), \ldots, G_N(s) \},
    \end{equation}
    where each $G_i(s)$ ($i = 1, 2, \ldots, N$) denotes an agent, e.g., a converter. The network coupling matrix $H(s) = \widetilde{\bf Y}^{-1}_{\rm grid}(s)$ is constant and positive definite. On this basis, we develop the following geometric decentralized stability condition.

    \begin{theorem}[{\bf Decentralized Stability Condition}]
    \label{thm:Decentralized}
    \,\\
    Consider $G$, $H \in \mathcal{RH}_{\infty}^{m\times m}$, where $G(s)$ is block-diagonal as in~\eqref{eq:diag} and $H(s) = H \succ 0$.
    The closed-loop system \( G(s)\#H(s) \) is stable if, for each \( \omega \in [0,\infty) \),
    \textbf{either}
    \begin{enumerate}[1)]
        \item \label{theGain}
        the decentralized \textbf{gain} condition, i.e.,
        \begin{equation}\label{eq:DecGain}
        \max_i \, \overline{\sigma}(G_i(j\omega)) 
        < \underline{\sigma}(H^{-1})
        \  \quad{\rm \textit{holds, \textbf{or}}}
        \end{equation}
        
        \item \label{thePhase}
        the decentralized \textbf{phase} condition, i.e.,
        \begin{equation}\label{eq:DecPhase}
        \begin{cases}
        \textit{a) } \max\limits_i \, \overline{\phi}(G_i(j\omega)) < \pi, \\[2pt]
        \textit{b) } \min\limits_i \, \underline{\phi}(G_i(j\omega)) > -\pi, \\[2pt]
        \textit{and c) } \max\limits_i \, \overline{\phi}(G_i(j\omega)) 
        - \min\limits_i \, \underline{\phi}(G_i(j\omega)) < \pi,
        \end{cases}
        \vspace{-2mm}
        \end{equation}
        \textit{holds, \textbf{or}}
        
        \item \label{theXZ}
        the decentralized \textbf{\(\bm{x}\)-\(\bm{z}\) graph separation}, i.e.,
        \begin{equation}\label{eq:DecXZ}
        \forall i: \,
        P(G_i(j\omega)) \cap 
        P\!\left(-\tfrac{1}{\tau}H^{-1}\right) = \varnothing,
        \,\forall \tau \in (0,1]
        \vspace{0mm}
        \end{equation}
        \textit{holds.}\vspace{-2mm}
    \end{enumerate}
    \end{theorem}

    \begin{pf}
    According to Corollary~IV.2 in~\citep{Huang2024}, 
    Eqs.~\eqref{eq:DecGain} and~\eqref{eq:DecPhase} 
    are equivalent to the small-gain and small-phase conditions in Theorem~\ref{prop:FW}. 
    We next show that condition~\eqref{eq:DecXZ}, though decentralized, leads to condition~\ref{cond:xz} in Theorem~\ref{prop:FW}. 

    As a first step, we discuss how $DW(-\tfrac{1}{\tau}H^{-1})$ extends under $\tau \in (0,1]$.
    When $\tau = 1$, $DW(-\tfrac{1}{\tau}H^{-1})$ is a polygon where all the vertices lies on $z=x^2, y=0$ because $H \succ 0$, as exemplified in Fig.~\ref{Fig_DW_H}.
    The trajectory of this DW shell with $\tau \in (0,1]$ lies on the left half $x$-$z$ plane ($y=0$), illustrated by the red area in Fig.~\ref{Fig_DW_H}. 
    The lower boundary of this trajectory is part of the \textit{parabola} $z = x^2$ and $x \in (-\infty,-\lambda_1(H^{-1})]$. The upper boundary consists of part of the side of the polygon and part of the parabola $z = a x^2$, which is concave upward, as shown in Fig.~\ref{Fig_DW_H}.

    \begin{figure}[!t]
    	\centering
    	\includegraphics[width=3.3in]{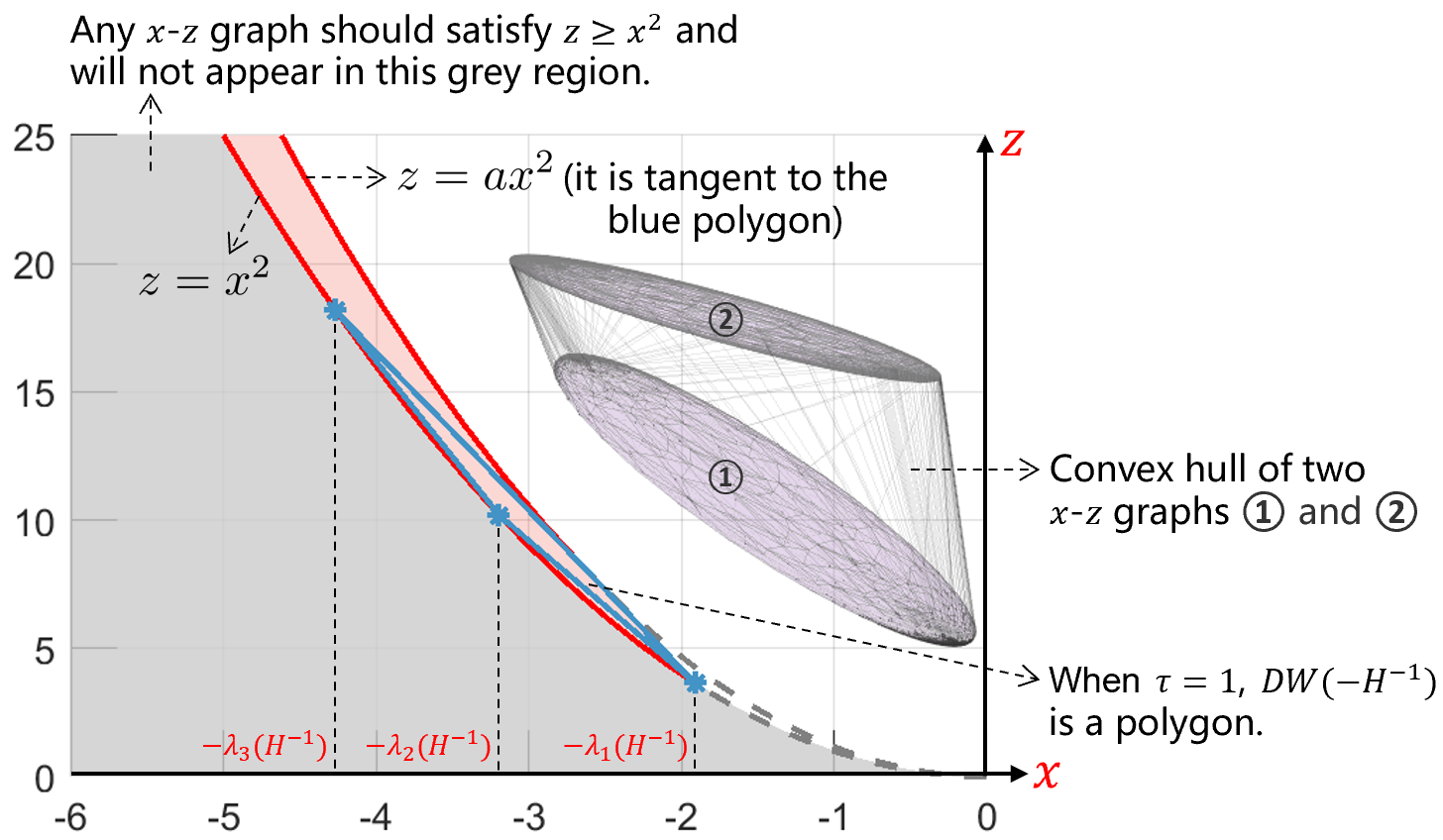}
    	\vspace{-2mm}
    	%\DeclareGraphicsExtensions.
    	\caption{Illustration of the DW shell trajectory of $-\tfrac{1}{\tau}H^{-1}$ with $\tau \in (0,1]$ when $H$ is 3-by-3 (i.e., the red area). The eigenvalues of a positive definite matrix $A \in \mathbb{R}^{n \times n}$ are organized as $\lambda_1(A) \le \lambda_2(A) \le \dots \le \lambda_n(A)$.}
    	\vspace{-1mm}
    	\label{Fig_DW_H}
    \end{figure}
    
    For the block-diagonal matrix $G(j\omega)=\mathrm{diag}\{G_i(j\omega)\}$, its DW shell satisfies $DW(G(j\omega))=\mathrm{conv}\big(\!\cup_i DW(G_i(j\omega))\big)$, 
    and the projection preserves convexity, so $P(G(j\omega))=\mathrm{conv}\big(\!\cup_i P(G_i(j\omega))\big)$. In other words, the DW shell (or $x$-$z$ graph) of a block-diagonal matrix is the convex hull of the DW shells (or $x$-$z$ graphs) of all its blocks. 
    We give an example in Fig.~\ref{Fig_DW_H} to show how a convex hull looks like.
    According to~\citep{zhang2025phantom}, the DW shell of any matrix lies above or intersects with the paraboloid $z = x^2+y^2$, so the $x$-$z$ graph of any matrix belongs to $z \ge x^2$ and will not appear in the grey region in Fig.~\ref{Fig_DW_H}. 
    Hence, the $x$-$z$ graph of any matrix will intersect with the red area only through the upper boundary of the red area which is concave upward.
    Then, it can be further deduced that if condition~\eqref{eq:DecXZ} holds, that is, each $P(G_i(j\omega))$ does not intersect with the red area in Fig.~\ref{Fig_DW_H}, their convex hull $P(G(j\omega))$ will neither intersect with the red area because the upper boundary of this area is concave upward, that is, condition~\ref{cond:xz} in Theorem~\ref{prop:FW} holds. 
    
    Hence, for each $\omega\in[0,\infty)$, if condition~\eqref{eq:DecGain}, \eqref{eq:DecPhase}, or~\eqref{eq:DecXZ} holds, then condition 1, 2, or 4 of Theorem~\ref{prop:FW} holds, so $G(s)\#H(s)$ is stable. This completes the proof. \unskip\nobreak\hfill $\square$
    \end{pf}

    Theorem~\ref{thm:Decentralized} proposes a geometric decentralized stability condition for structured network systems. 
    Since conditions~\eqref{eq:DecGain}, \eqref{eq:DecPhase}, and \eqref{eq:DecXZ} focus on the characteristics of each $G_i(s)$, the overall system stability can be efficiently assessed in a decentralized manner by examining individual subsystems, 
    enabling scalable and modular analysis. Note that Theorem~\ref{thm:Decentralized} provides a sufficient stability condition, namely, the system is stable if at any frequency, one of the three conditions holds. 
    Unlike Theorem~\ref{prop:FW}, we do not employ DW shell separation or numerical range separation for decentralized analysis. The reason is that even if each $DW(G_i(j\omega))$ is separated from the trajectory of the network's DW shell, their convex hull $DW(G(j\omega))$ may still intersect with it in the 3D space. 
    A similar limitation also exists when applying the numerical range separation. By contrast, the positive definiteness of $H$ ensures that the upper bound of its $x$-$z$ graph trajectory (over $\tau \in (0,1]$) is concave upward, which allows decentralized verification of the $x$-$z$ graph separation, as discussed in the proof of Theorem~\ref{prop:FW}. 
    In practice, such $x$-$z$ graph separation serves as a complementary condition to the small-gain and small-phase conditions to reduce the conservativeness.

    \section{Applications of Geometric Decentralized Stability Condition in Power Systems}\label{sec:V}
    
    In this section, we apply the geometric decentralized stability condition to analyze multi-converter power systems. Firstly, the mixed geometric stability condition in Theorem~\ref{prop:FW} is applied to evaluate the stability of a single-converter system, which geometrically demonstrates how the converter interacts with the power grid. Then, the decentralized stability condition in Theorem~\ref{thm:Decentralized} is employed to assess the stability of a three-converter system, with an example illustrating how an appropriate change of the converter's control can improve the system stability.

    \subsection{Case Studies of a Single-Converter System}\label{single}
\begin{figure}[!t]
    	\centering
    	\includegraphics[width=2.6in]{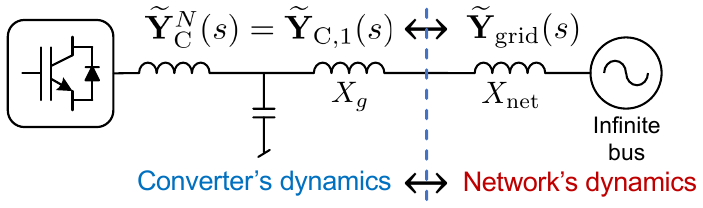}
    	\vspace{-3mm}
    	%\DeclareGraphicsExtensions.
    	\caption{A single converter connected to an infinite bus.}
    	\vspace{0mm}
    	\label{Fig_one_conv}
    \end{figure}

    Consider the single-converter system in Fig.~\ref{Fig_one_conv}, where the converter operates in grid-following (GFL) mode. The GFL control structure is the same as that in~\citep{Huang2024}, with the phase-locked loop (PLL) bandwidth being 40~rad/s. The converter dynamics are described by \(\widetilde{\bf Y}_{{\rm C}}^N(s)=\widetilde{\bf Y}_{{\rm C,1}}(s)\) as there is only one converter. The short-circuit ratio (SCR)~\citep{dong2018small} of the system is \({\rm SCR} = 1/X_{\rm net} = 1.6\), where \(X_{\rm net}\) denotes the line reactance, and the AC grid dynamics are represented by \(\widetilde{\bf Y}_{\rm grid}(s)=X_{\rm net}^{-1}I_2\). 
    Consider the closed-loop system in Fig.~\ref{fdbk}, and let \(G(s)=\widetilde{\bf Y}_{{\rm C,1}}(s)\) and \(H(s)=\widetilde{\bf Y}_{\rm grid}^{-1}(s)\). 
    The overall closed-loop dynamics can then be expressed as \(G(s)\#H(s)=\widetilde{\bf Y}_{{\rm C,1}}(s)\#\widetilde{\bf Y}_{\rm grid}^{-1}(s)\). 
    We use the condition in Theorem~\ref{prop:FW} to analyze the system stability.

    \begin{figure}[!t]
    	\centering
    	\includegraphics[width=3in]{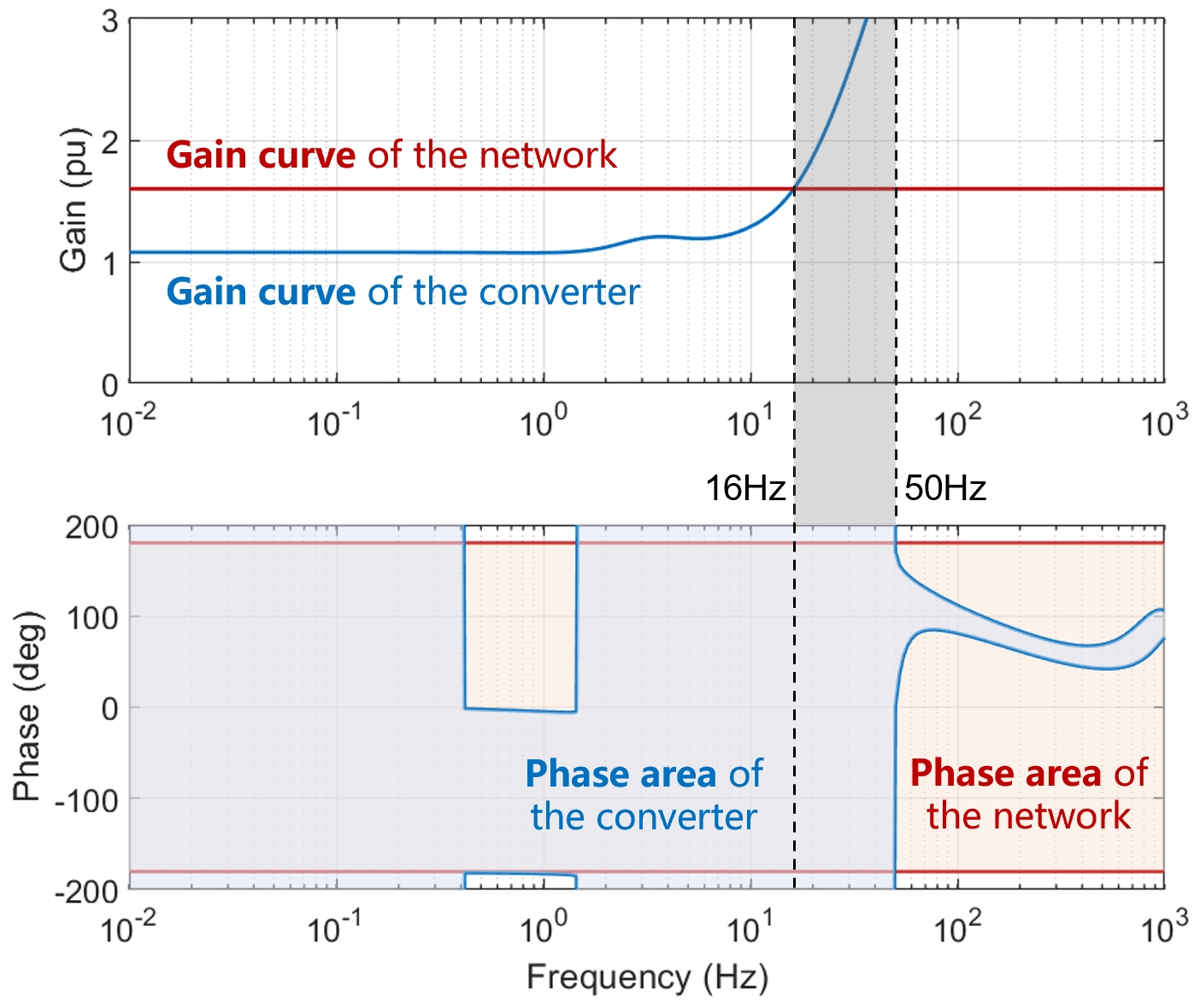}
    	\vspace{-1mm}
    	%\DeclareGraphicsExtensions.
    	\caption{Gain and phase of the single-converter system.}
    	\vspace{1mm}
    	\label{Fig_one_gainphase}
    \end{figure}

    \begin{figure}[!t]
    	\centering
    	\includegraphics[width=3.4in]{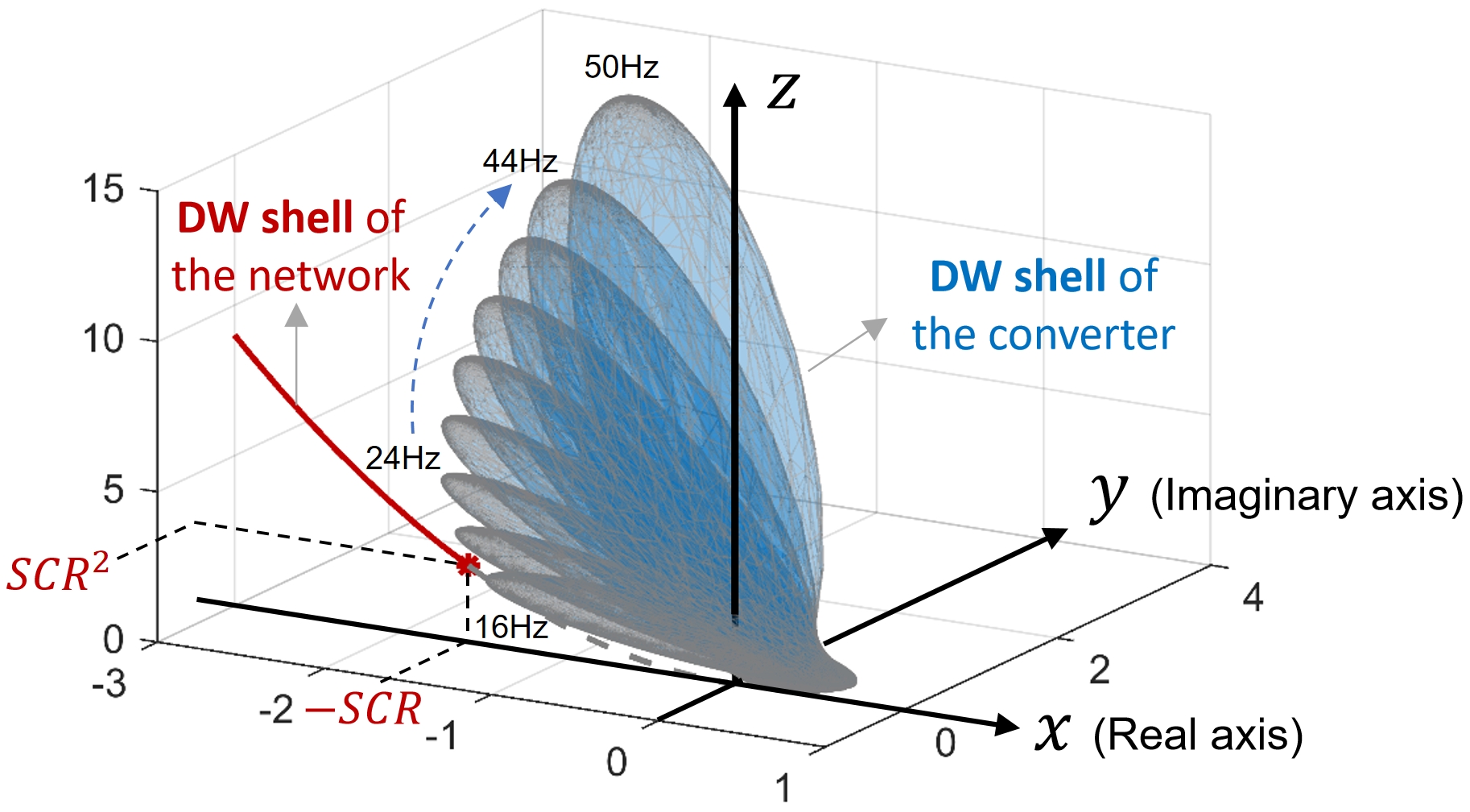}
    	\vspace{-2mm}
    	%\DeclareGraphicsExtensions.
    	\caption{DW shells of the converter at different frequencies (blue ellipsoids) and the DW shell trajectory (over $\tau \in (0,1]$) of the network (red line).}
    	\vspace{0mm}
    	\label{Fig_one_3DDW}
    \end{figure}

    \begin{figure}[!t]
    	\centering
    	\includegraphics[width=2.6in]{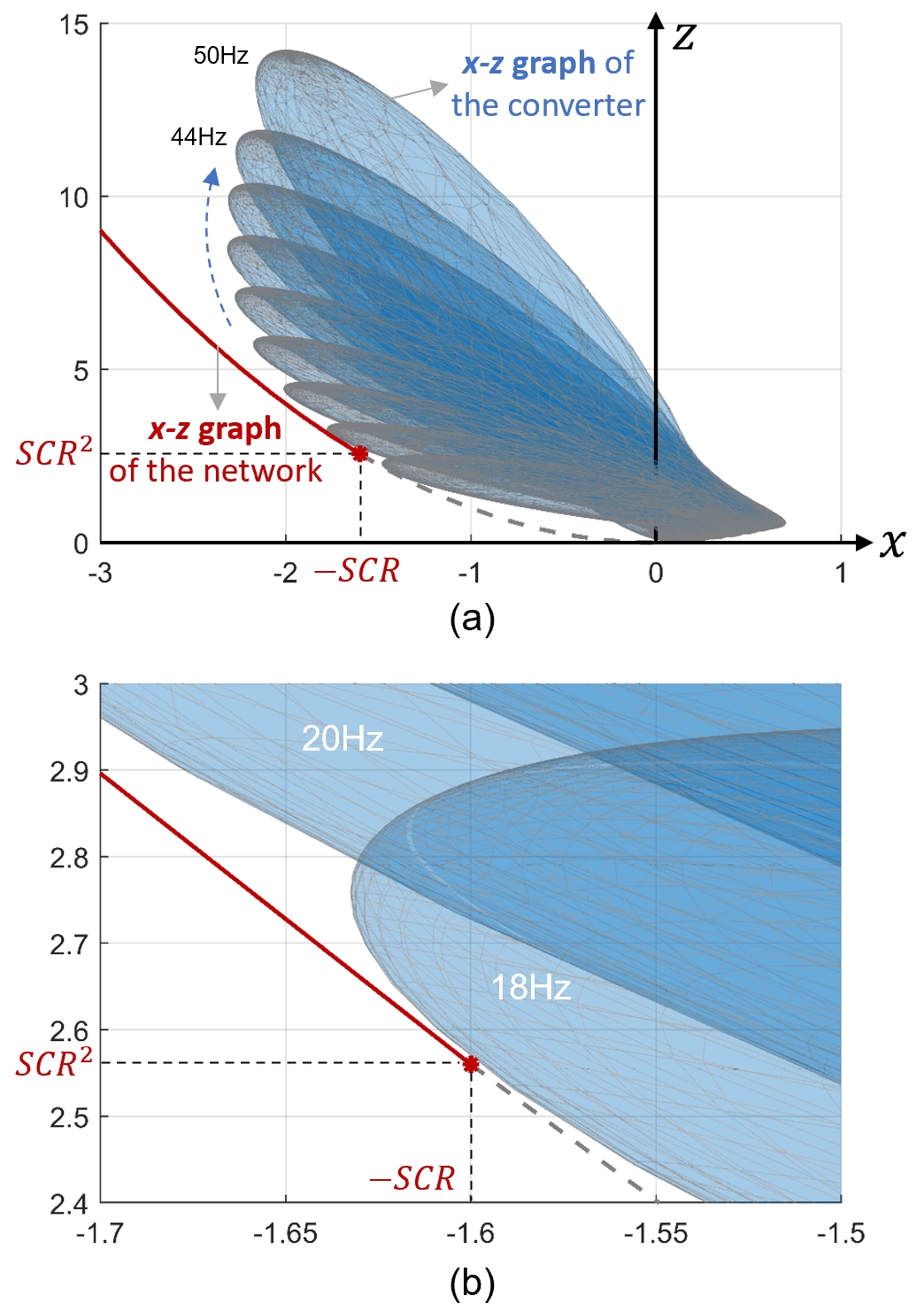}
    	\vspace{0mm}
    	%\DeclareGraphicsExtensions.
    	\caption{The converter's \(x\)-\(z\) graphs at different frequencies (blue ellipses) and the \(x\)-\(z\) graph trajectory (over $\tau \in (0,1]$) of the network (red line).}
    	\vspace{0mm}
    	\label{Fig_one_xzDW}
    \end{figure}

    The phase areas of the converter and the network are respectively defined by $[\underline{\phi}(\widetilde{\bf Y}_{{\rm C,1}}(j\omega)),\,\overline{\phi}(\widetilde{\bf Y}_{{\rm C,1}}(j\omega))]$ and $[-\pi-\underline{\phi}(\widetilde{\bf Y}_{\rm grid}^{-1}(j\omega)),\ \pi-\overline{\phi}(\widetilde{\bf Y}_{\rm grid}^{-1}(j\omega))]$. The gain curves of the converter and the network are respectively defined by $\overline{\sigma}(\widetilde{\bf Y}_{{\rm C,1}}(j\omega))$ and $\underline{\sigma}(\widetilde{\bf Y}_{\rm grid}(j\omega))$.
    According to~\citep{Huang2024}, if the converter's gain is smaller than the network's gain, then the small-gain condition holds; if the converter's phase area is contained in the network's phase area, then the small-phase condition holds. Fig.~\ref{Fig_one_gainphase} plots the gain curves and phase areas of the system. 
    Below 16~Hz, the converter's gain is smaller than the network's gain (i.e., $1/X_{\rm net}={\rm SCR}=1.6$), thus satisfying the small-gain condition. Above 50~Hz, the converter’s phase area is contained in the network’s phase area (i.e., $[-\pi,\pi]$), thus satisfying the small-phase condition. 
    
    However, between 16~Hz and 50~Hz, neither the small-gain nor the small-phase condition holds. In this frequency range, one can further check conditions 3, 4, or 5 in Theorem~\ref{prop:FW}. Note that condition~5 (DW shell separation) is the less conservative one.
    Fig.~\ref{Fig_one_3DDW} plots the DW shells of the converter and the network from 16~Hz to 50~Hz. For the converter, the DW shells are obtained from $\widetilde{\bf Y}_{{\rm C},1}(j\omega)$ at different frequencies, where each DW shell is an ellipsoid. For the network, since we have $\widetilde{\bf Y}_{\rm grid}(j\omega)=X_{\rm net}^{-1}I_2$ which is positive definite, the DW shell trajectory of $-\tfrac{1}{\tau}\widetilde{\bf Y}_{\rm grid}(j\omega)$ ($\tau\in(0,1]$) is part of a parabola on the \(x\)-\(z\) plane, i.e., $z=x^2$ with $y=0$, as shown by the red line in Fig.~\ref{Fig_one_3DDW}. The $x$-axis range of this red line is $(-\infty,-{\rm SCR}]$. It can be seen that the converter’s DW shells are separated from the network’s DW shell, satisfying condition~5 in Theorem~\ref{prop:FW}. Hence, the system is stable.

    According to~\citep{zhang2025phantom}, the DW shell of any complex matrix lies above or intersects with the paraboloid $z = x^2 + y^2$. As shown in Fig.~\ref{Fig_one_3DDW}, the network’s DW shell (the red line) lies exactly on this paraboloid with $y = 0$ and $z = x^2$. Hence, for the single-converter system, condition~4 is equivalent to condition~5 in Theorem~\ref{prop:FW}, and the system stability can be analyzed via the 2D \(x\)-\(z\) graphs, which are more intuitive and convenient than the 3D DW shells.  
    Fig.~\ref{Fig_one_xzDW} plots the \(x\)-\(z\) graphs of the converter and network, where (b) is the zoom-in version of (a). It can be observed that the converter’s \(x\)-\(z\) graph remains separated from the network’s \(x\)-\(z\) graph (part of the parabola $z = x^2$) from 16~Hz to 50~Hz. At 18~Hz, the converter’s \(x\)-\(z\) graph is very close to the network's \(x\)-\(z\) graph but remains separated, which indicates a small stability margin.

    To sum up, the single-converter system satisfies: 1) the small-gain condition below 16~Hz; 2) the small-phase condition above 50~Hz; and 3) the DW shell separation (or equivalently, the \(x\)-\(z\) graph separation) condition between 16~Hz and 50~Hz. Hence, the closed-loop system is stable according to Theorems~\ref{prop:FW} or~\ref{thm:Decentralized}. Note that the DW shell separation is equivalent to the $x$-$z$ graph separation only in the single-converter case. In multi-converter systems, they are not equivalent, and only the $x$-$z$ graph separation enables decentralized stability analysis.

    \vspace{-2mm}

    \subsection{Case Studies of a Three-Converter System}\label{three}
    \vspace{-2mm}
    
\begin{figure}[!t]
    	\centering
        \vspace{0.5mm}
    	\includegraphics[width=3.2in]{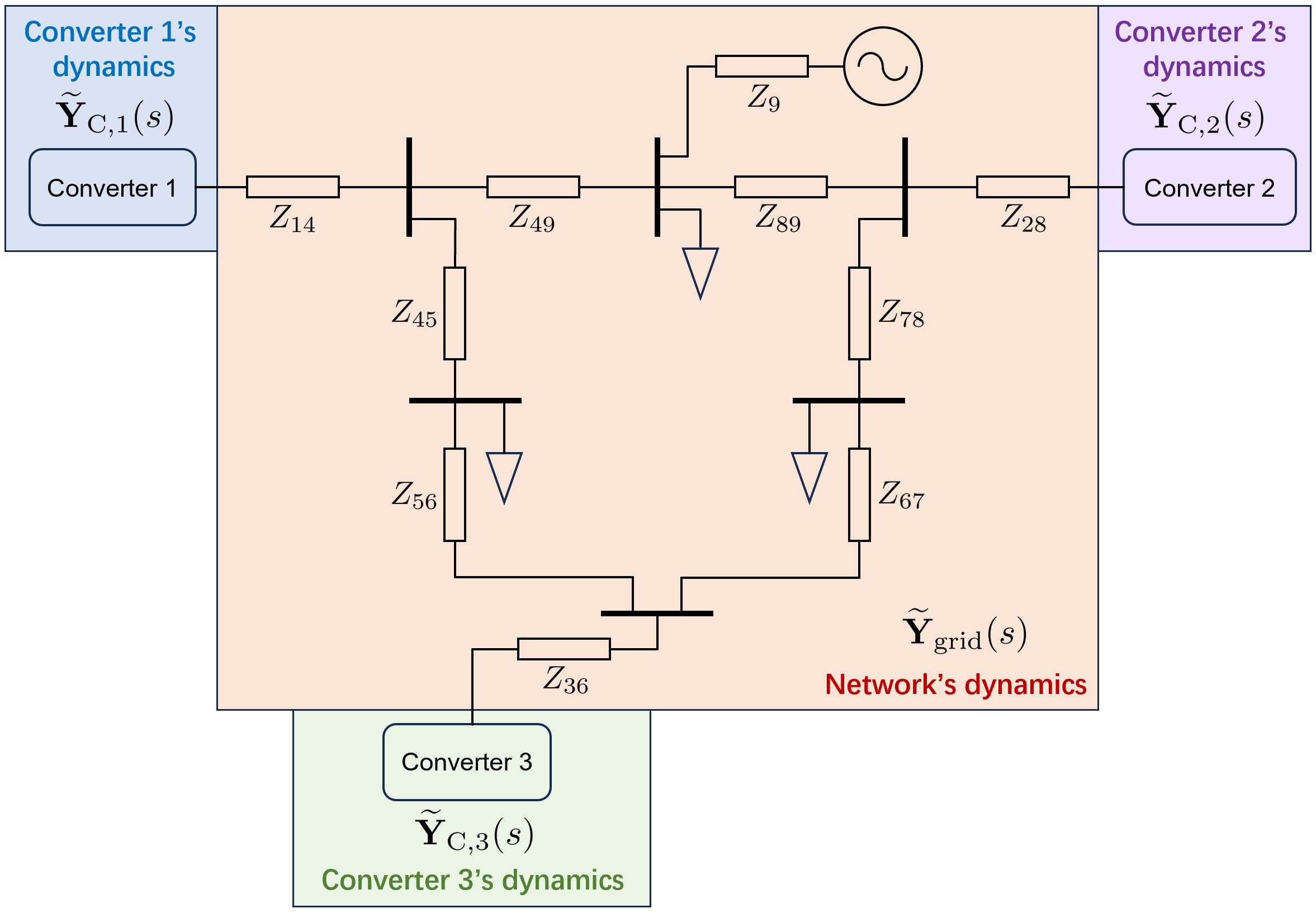}
    	\vspace{-2mm}
    	%\DeclareGraphicsExtensions.
    	\caption{A three-converter test system.}
    	\vspace{0mm}
    	\label{Fig_three_converter}
    \end{figure}

    We consider now the three-converter system in Fig.~\ref{Fig_three_converter}, where all the converters operate in GFL mode. The PLL bandwidths of the three converters are 40~rad/s, 20~rad/s, and 95~rad/s, respectively. According to Section~III, the converters' dynamics can be represented by the block-diagonal matrix $\widetilde{\bf Y}_{{\rm C}}^N(s) = \mathrm{diag}\{\widetilde{\bf Y}_{{\rm C},1}(s), \widetilde{\bf Y}_{{\rm C},2}(s), \widetilde{\bf Y}_{{\rm C},3}(s)\}$, and the network dynamics are given by $\widetilde{\bf Y}_{\rm grid}(s) = {\bf S}^{-\frac{1}{2}}{\bf B}_{\rm r}{\bf S}^{-\frac{1}{2}} \otimes I_2$, which is a positive definite constant matrix. We again let $G(s) = \widetilde{\bf Y}_{{\rm C}}^N(s)$ and $H(s) = \widetilde{\bf Y}_{\rm grid}^{-1}(s)$ and analyze system stability using Theorem~\ref{thm:Decentralized}.

\begin{figure}[!t]
    	\centering
    	\includegraphics[width=3in]{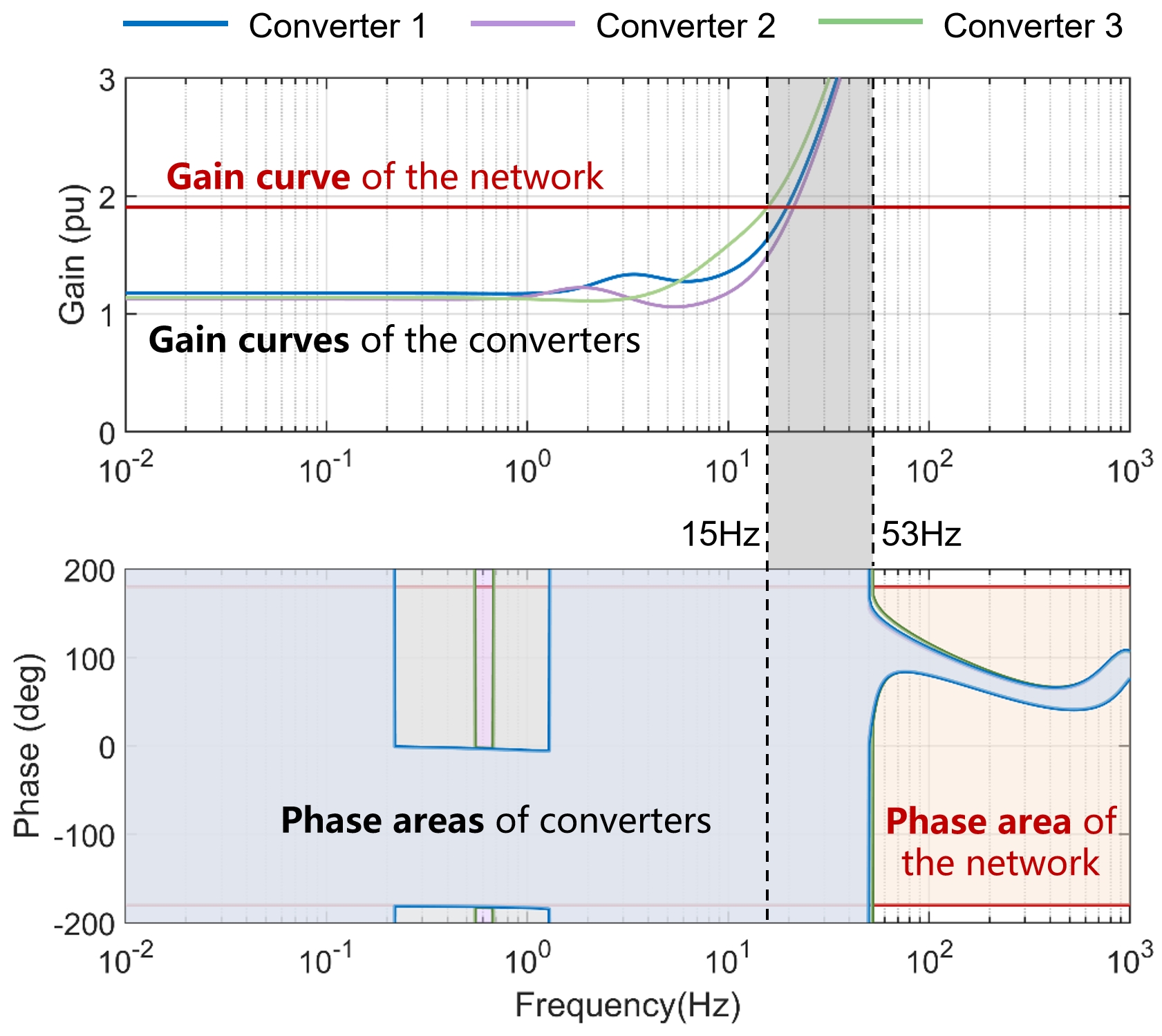}
    	\vspace{-2mm}
    	%\DeclareGraphicsExtensions.
    	\caption{Gain and phase of the three-converter system.}
    	\vspace{0mm}
    	\label{Fig_three_gainphase}
    \end{figure}

    Fig.~\ref{Fig_three_gainphase} plots the gain curves and phase areas of the three converters and the network, which are determined by $\widetilde{\bf Y}_{{\rm C},i}(j\omega)$ and $\widetilde{\bf Y}_{\rm grid}(j\omega)$, respectively. Below 15~Hz, the gains of all the converters $\overline{\sigma}(\widetilde{\bf Y}_{{\rm C,i}}(j\omega))$ are smaller than that of the network $\underline{\sigma}(\widetilde{\bf Y}_{\rm grid}(j\omega))$, satisfying the decentralized gain condition in Theorem~\ref{thm:Decentralized}.
    We notice that the gain of the network is in fact $\underline{\sigma}(\widetilde{\bf Y}_{\rm grid}(j\omega))=\underline{\sigma}({\bf S}^{-\frac{1}{2}}{\bf B}_{\rm r}{\bf S}^{-\frac{1}{2}})={\rm gSCR}$, where gSCR denotes the so-called generalized short-circuit ratio in power systems~\citep{dong2018small, xin2022many}. Above 53~Hz, the phase areas of all the converters (the largest phase difference is less than $180^\circ$) are contained in the phase area of the network (i.e., $[-\pi,\pi]$), satisfying the decentralized phase condition in Theorem~\ref{thm:Decentralized}. Between 15~Hz and 53~Hz, neither the gain condition nor the phase condition holds, and according to Theorem~\ref{thm:Decentralized}, the system stability can be further analyzed by checking condition~\ref{theXZ} through \(x\)-\(z\) graphs.

    \begin{figure*}[!t]
     \vspace{0mm}
    \centerline{\includegraphics[width=0.87\linewidth]{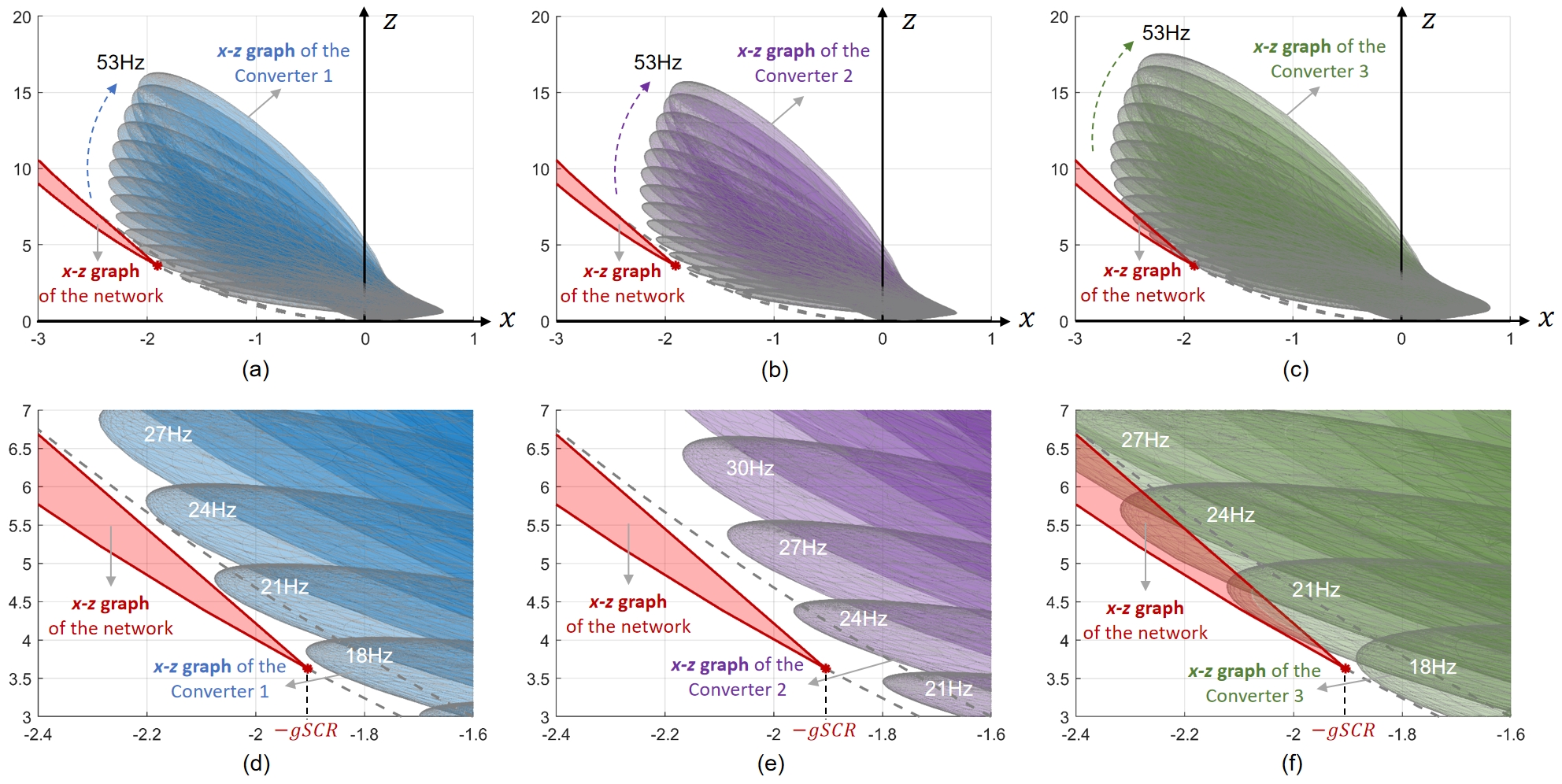}}
    	\vspace{-3mm}
            \caption{The \(x\)-\(z\) graphs of the converters and the \(x\)-\(z\) graph trajectory of the network. (a) \& (d): Converter~1. (b) \& (e): Converter~2. (c) \& (f): Converter~3. Subplots (d, e, f) are respectively the zoom-in versions of (a, b, c).}
            \label{Fig_three_xz}
    	\vspace{1mm}
    \end{figure*}

    \begin{figure}[!t]
    	\centering
    	\includegraphics[width=2.6in]{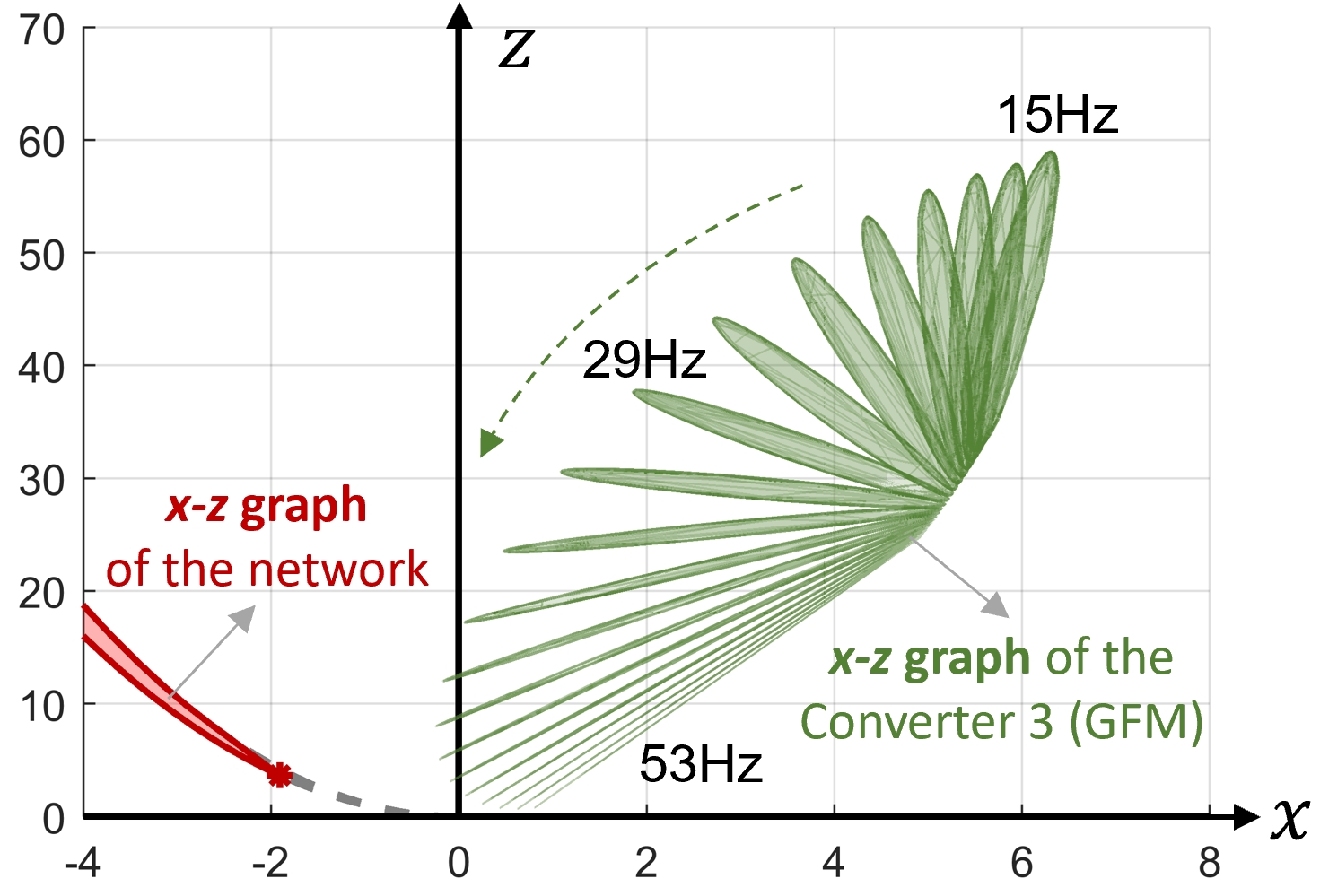}
    	\vspace{-2mm}
    	%\DeclareGraphicsExtensions.
    	\caption{The \(x\)-\(z\) graphs of Converter 3 (with GFM control) and the \(x\)-\(z\) graph trajectory of the network.}
    	\vspace{0mm}
    	\label{Fig_three_GFM_xz}
    \end{figure}

\begin{figure}[!t]
    	\centering
        \vspace{0mm}
    	\includegraphics[width=2.4in]{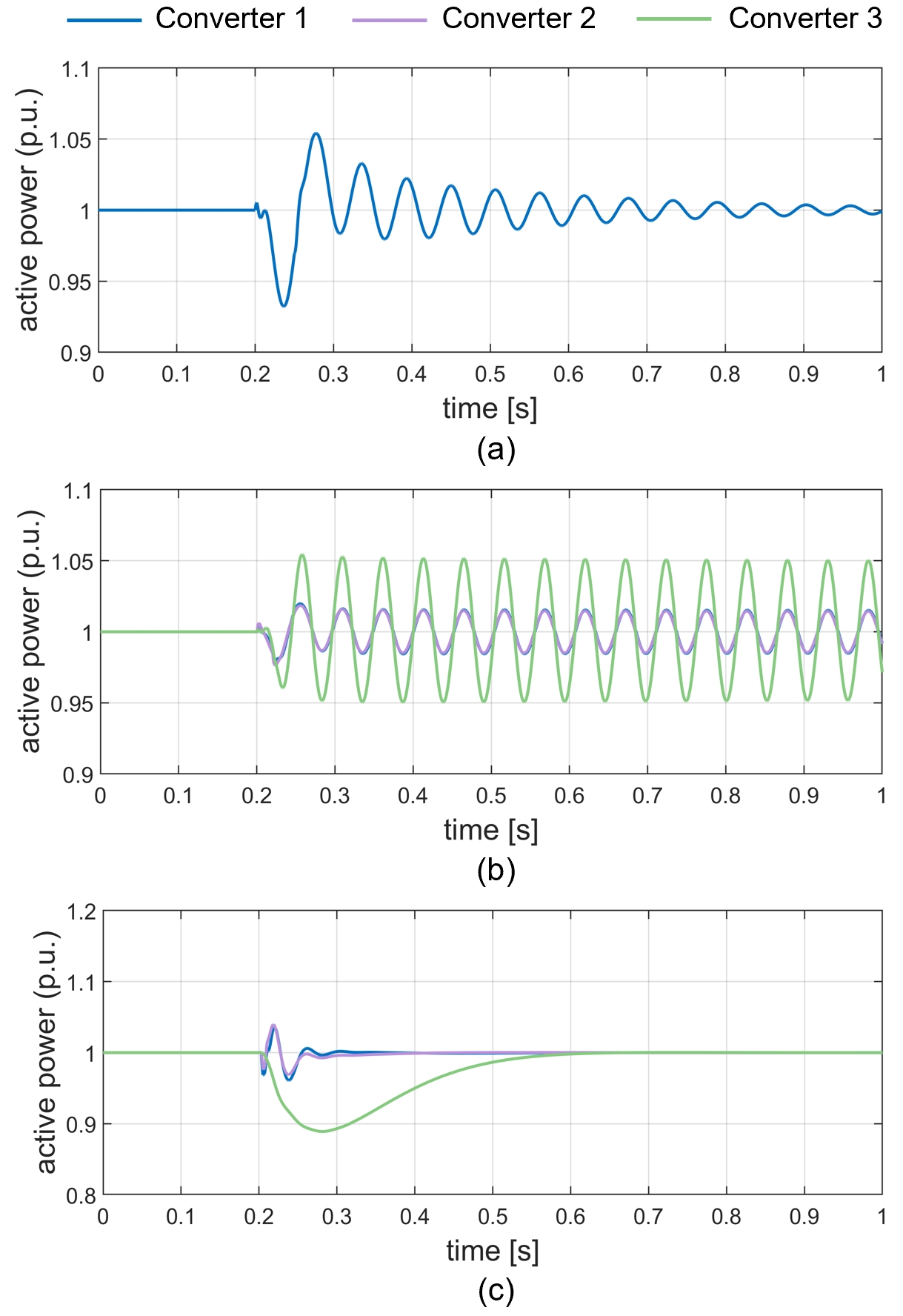}
    	\vspace{-3mm}
    	%\DeclareGraphicsExtensions.
    	\caption{Time-domain responses of the test systems: (a) single-converter system; (b) three-converter system (all GFL); and (c) modified three-converter system (Converters~1-2: GFL; Converter~3: GFM).}
    	\vspace{1mm}
    	\label{Fig_three_results}
    \end{figure}

    Fig.~\ref{Fig_three_xz} plots the \(x\)-\(z\) graphs of the three converters and the network from 15~Hz to 53~Hz, where the \(x\)-\(z\) graph trajectory of the network (over $\tau \in (0,1]$) constitutes the red region (its upper boundary is upward concave and its lower boundary corresponds to a part of the parabola \(z = x^2\) with \(x \in (-\infty, -{\rm gSCR}]\)). It can be observed from Fig.~\ref{Fig_three_xz}~(a) and (d) that the \(x\)-\(z\) graphs of Converter~1 are separated from the red region (i.e., the \(x\)-\(z\) graph trajectory of the network). 
    Similarly, as shown in Fig.~\ref{Fig_three_xz}~(b) and (e), the \(x\)-\(z\) graphs of Converter~2 are also separated from the red region. However, in Fig.~\ref{Fig_three_xz}~(c) and (f), the \(x\)-\(z\) graph of Converter~3 intersects with the red region, indicating that condition~\ref{theXZ} in Theorem~\ref{thm:Decentralized} is violated. Therefore, the system stability cannot be guaranteed in this frequency range, suggesting a potential instability risk induced by Converter~3. This apparent risk is confirmed in the simulations in Fig.~\ref{Fig_three_results}~(b) showing sustained oscillations.

    To improve the stability, Converter~3 is then replaced by a grid-forming (GFM) converter, which generally has better dynamic performance~\citep{Huang2024}. 
    Converters~1 and~2 remain unchanged as in Fig.~\ref{Fig_three_xz}~(a) and (b). The resulting \(x\)-\(z\) graphs of Converter~3 (GFM) are plotted in Fig.~\ref{Fig_three_GFM_xz}, showing that the converter’s \(x\)-\(z\) graphs are clearly separated from the red region. According to Theorem~\ref{thm:Decentralized}, this \(x\)-\(z\) graph separation confirms that the modified three-converter system is stable.

    \subsection{Time-Domain Simulation Results}
    \vspace{-2mm}
    We next provide simulation results of the test systems. Fig.~\ref{Fig_three_results}~(a) shows the response of the single-converter system in Fig.~\ref{Fig_one_conv}, where the grid voltage experiences a 5\% voltage sag at $t = 0.2~\mathrm{s}$ and then recovers. 
    The system is stable but exhibits weakly-damped oscillations, which is consistent with Fig.~\ref{Fig_one_xzDW} where the converter’s $x$-$z$ graph at 18~Hz is very close to that of the network (red line).
    Fig.~\ref{Fig_three_results}~(b) shows the simulation results of the three-converter system in Fig.~\ref{Fig_three_converter} under the same disturbance. When all converters operate in GFL mode, the system is marginally stable and exhibits sustained oscillations, aligned with Fig.~\ref{Fig_three_xz}~(c) and (f) where Converter~3's $x$-$z$ graphs intersect with the red region. Fig.~\ref{Fig_three_results}~(c) shows the system response when Converter~3 operates in GFM mode. The system is stable and has a satisfactory damping ratio, consistent with Fig.~\ref{Fig_three_GFM_xz} where the \(x\)-\(z\) graphs of Converter~3 (GFM) are separated from the red region. These simulation results confirm the effectiveness of the geometric decentralized stability analysis.

    \section{Conclusions}
    \vspace{-2mm}
    This paper presents a geometric decentralized stability analysis framework based on the concept of gain, phase, DW shell, and its projection (\(x\)-\(z\) graph). The proposed condition provides a geometric interpretation of the small-gain theorem and the small-phase theorem, and further introduces the condition of \(x\)-\(z\) graph separation to reduce the conservativeness. Our approach enables decentralized stability analysis of multi-agent network systems in a scalable and modular manner. Applied to multi-converter power systems, our approach offers an intuitive geometric understanding of the converter-grid interactions and facilitates the modular and scalable stability analysis for large-scale power systems. Future work will consider more sophisticated components such as shunt capacitors in the network and investigate the control synthesis problem via the gain, phase, and \(x\)-\(z\) graph conditions.

    \normalem
	\bibliography{references}

\begin{thebibliography}{13}
\providecommand{\natexlab}[1]{#1}
\providecommand{\url}[1]{\texttt{#1}}
\providecommand{\urlprefix}{URL }
\expandafter\ifx\csname urlstyle\endcsname\relax
  \providecommand{\doi}[1]{doi:\discretionary{}{}{}#1}\else
  \providecommand{\doi}{doi:\discretionary{}{}{}\begingroup
  \urlstyle{rm}\Url}\fi

\bibitem[{Chen et~al.(2024)Chen, Wang, Khong, and Qiu}]{chen2024phase}
Chen, W., Wang, D., Khong, S.Z., and Qiu, L. (2024).
\newblock A phase theory of multi-input multi-output linear time-invariant
  systems.
\newblock \emph{SIAM Journal on Control and Optimization}, 62(2), 1235--1260.

\bibitem[{Davis(1968)}]{davis1968shell}
Davis, C. (1968).
\newblock The shell of a {Hilbert-space} operator.
\newblock \emph{Acta Sci. Math.(Szeged)}, 29(1-2), 69--86.

\bibitem[{Dong et~al.(2018)Dong, Xin, Wu, and Huang}]{dong2018small}
Dong, W., Xin, H., Wu, D., and Huang, L. (2018).
\newblock Small signal stability analysis of multi-infeed power electronic
  systems based on grid strength assessment.
\newblock \emph{IEEE trans. Power Systems}, 34(2).

\bibitem[{Huang et~al.(2024)Huang, Wang, Wang, Xin, Ju, Johansson, and
  Dörfler}]{Huang2024}
Huang, L., Wang, D., Wang, X., Xin, H., Ju, P., Johansson, K.H., and Dörfler,
  F. (2024).
\newblock Gain and phase: Decentralized stability conditions for power
  electronics-dominated power systems.
\newblock \emph{IEEE Trans. Power Systems}, 39(6), 7240--7256.
\newblock \doi{10.1109/TPWRS.2024.3380528}.

\bibitem[{Huang et~al.(2020)Huang, Xin, and D{\"o}rfler}]{huang2020h}
Huang, L., Xin, H., and D{\"o}rfler, F. (2020).
\newblock H$\infty$-control of grid-connected converters: Design, objectives
  and decentralized stability certificates.
\newblock \emph{IEEE Trans. Smart Grid}, 11(5), 3805--3816.

\bibitem[{Lestas(2012)}]{lestas_DW}
Lestas, I. (2012).
\newblock Large scale heterogeneous networks, the {D}avis--{W}ielandt shell,
  and graph separation.
\newblock \emph{SIAM Journal on Control and Optimization}, 50(4), 1753--1774.

\bibitem[{Li et~al.(2008)Li, Poon, and Sze}]{Li_DW}
Li, C.K., Poon, Y.T., and Sze, N.S. (2008).
\newblock Eigenvalues of the sum of matrices from unitary similarity orbits.
\newblock \emph{SIAM Journal on Matrix Analysis and Applications}, 30(2),
  560--581.

\bibitem[{Milano et~al.(2018)Milano, D{\"o}rfler, Hug, Hill, and
  Verbi{\v{c}}}]{milano2018foundations}
Milano, F., D{\"o}rfler, F., Hug, G., Hill, D.J., and Verbi{\v{c}}, G. (2018).
\newblock Foundations and challenges of low-inertia systems.
\newblock In \emph{2018 power systems computation conference (PSCC)}, 1--25.
  IEEE.

\bibitem[{Wang et~al.(2020)Wang, Chen, Khong, and Qiu}]{wang2020phases}
Wang, D., Chen, W., Khong, S.Z., and Qiu, L. (2020).
\newblock On the phases of a complex matrix.
\newblock \emph{Linear Algebra and its Applications}, 593, 152--179.

\bibitem[{Wielandt et~al.(1955)}]{wielandt1955eigenvalues}
Wielandt, H. et~al. (1955).
\newblock On eigenvalues of sums of normal matrices.
\newblock \emph{Pacific J. Math}, 5(4), 633--638.

\bibitem[{Xin et~al.(2024)}]{xin2022many}
Xin, H. et~al. (2024).
\newblock How many grid-forming converters do we need? a perspective from small
  signal stability and power grid strength.
\newblock \emph{IEEE Trans. Power Systems}, 40(1), 623--635.

\bibitem[{Zames(2003)}]{zames2003input}
Zames, G. (2003).
\newblock On the input-output stability of time-varying nonlinear feedback
  systems part one: Conditions derived using concepts of loop gain, conicity,
  and positivity.
\newblock \emph{IEEE trans. automatic control}, 11(2), 228--238.

\bibitem[{Zhang et~al.(2025)Zhang, Yang, Ringh, and Qiu}]{zhang2025phantom}
Zhang, D., Yang, X., Ringh, A., and Qiu, L. (2025).
\newblock The phantom of {D}avis-{W}ielandt shell: A unified framework for
  graphical stability analysis of {MIMO LTI} systems.
\newblock \emph{arXiv preprint arXiv:2507.19918}.

\end{thebibliography}

\end{document}